\documentstyle[12pt]{article}
\newcommand{\bea}{\begin{eqnarray}}
\newcommand{\eea}{\end{eqnarray}}
\newcommand{\beq}{\begin{equation}}
\newcommand{\eeq}{\end{equation}}

\def\misp{\slash\hspace{-5pt}p}
\def\msbar{\ifmmode{\overline{\rm MS}} \else{$\overline{\rm MS}$} \fi} 
\def\drbar{\ifmmode{\overline{\rm DR}} \else{$\overline{\rm DR}$} \fi} 
\def\st{\ifmmode{\tilde{t}} \else{$\tilde{t}$} \fi} 
\def\sb{\ifmmode{\tilde{b}} \else{$\tilde{b}$} \fi} 
\def\sq{\ifmmode{\tilde{q}} \else{$\tilde{q}$} \fi} 
\def\sg{\ifmmode{\tilde{g}} \else{$\tilde{g}$} \fi} 
\def\sz{\ifmmode{\tilde{\chi}^0} \else{$\tilde{\chi}^0$} \fi} 
\def\sw{\ifmmode{\tilde{\chi}} \else{$\tilde{\chi}$} \fi} 
\def\sl{\ifmmode{\tilde{\ell}} \else{$\tilde{\ell}$} \fi} 
\def\sn{\ifmmode{\tilde{\nu}} \else{$\tilde{\nu}$} \fi} 
\def\stau{\ifmmode{\tilde{\tau}} \else{$\tilde{\tau}$} \fi} 
\def\nle{\rlap{\lower 3.5 pt \hbox{$\mathchar \sim$}}\raise 1pt \hbox{$<$}}
\def\nge{\rlap{\lower 3.5 pt \hbox{$\mathchar \sim$}}\raise 1pt \hbox{$>$}}

\begin{document}
\pagestyle{empty}
\vspace*{-3cm}
\begin{flushright}
UWThPh-1996-44\\
HEPHY-PUB 650/96\\
TGU-18\\
ITP-SU-96/03\\
 MADPH-96-952\\
hep-ph/9607388\\
\vspace{0.3cm}
July, 1996
\end{flushright}

\vspace{2cm}
\begin{center}
\begin{Large} \bf
Impact of squark pair decay modes on search for 
supersymmetric neutral Higgs bosons
\end{Large}
\end{center}
\vspace{10mm}

\begin{center}
\large A.~Bartl,$^1$ H.~Eberl,$^2$ K.~Hidaka,$^3$ T.~Kon,$^4$\\ 
W.~Majerotto$^2$ and Y.~Yamada$^5$
\end{center}
\vspace{0mm}

\begin{center}
\begin{tabular}{l}
$^1${\it Institut f\"ur Theoretische Physik, Universit\"at Wien, 
A-1090 Vienna, Austria}\\
$^2${\it Institut f\"ur Hochenergiephysik der \"Osterreichischen 
Akademie der Wissenschaften,}\\ 
{\it A-1050 Vienna, Austria}\\ 
$^3${\it Department of Physics, Tokyo Gakugei University, Koganei, 
Tokyo 184, Japan}\\
$^4${\it Faculty of Engineering, Seikei University, Musashino, 
Tokyo 180, Japan}\\
$^5${\it Department of Physics, University of Wisconsin, Madison, 
WI 53706, USA}\\ 
\end{tabular}
\end{center}
\vspace{10mm}

\begin{abstract}
\begin{small}
\baselineskip=28pt
We present a detailed study of the decays of the neutral Higgs bosons 
$H^0$ and $A^0$ within the Minimal Supersymmetric Standard Model. 
We find that the supersymmetric modes $\st\bar{\st}$ and 
$\sb\bar{\sb}$ can dominate the $H^0$ and $A^0$ decays in a wide 
range of the model parameters due to large Yukawa couplings and 
mixings of \st and \sb. 
Compared to the conventional modes $t\bar{t}$ and $b\bar{b}$, 
these modes have very distinctive signatures. 
This could have a decisive impact on the neutral Higgs boson 
searches at future colliders. 
\end{small}
\end{abstract}


\newpage
\pagestyle{plain}
\setcounter{page}{2}
\baselineskip=28pt

The existence of two or more Higgs bosons would be a clear indication 
that the Standard Model must be extended. 
For example, the Minimal Supersymmetric Standard Model (MSSM) 
\cite{1} with two Higgs doublets predicts the existence of 
five physical Higgs bosons $h^0$, $H^0$, $A^0$, and $H^\pm$ \cite{2,3}. 
In searching for the Higgs bosons it is necessary to study not only 
the production mechanism, but also all their possible decay modes. 
Their decays to supersymmetric (SUSY) particles could be very 
important if they are kinematically allowed. 
This is indeed the case for the charged Higgs boson $H^+$. 
If all SUSY particles are very heavy, the $H^+$ decays dominantly 
into $t\bar{b}$ ; 
the decays $H^+ \to \tau^+ \nu$ and/or $H^+ \to W^+ h^0$ are 
dominant below the $t\bar{b}$ threshold \cite{2,4}. 
In refs.\cite{5,6} all decay modes of $H^+$ including the 
SUSY-particle modes were studied in the the case that the 
SUSY-particles are relatively light ; it was shown that the SUSY 
decay mode $H^+ \to \st\bar{\sb}$ can be dominant in a large region 
of the MSSM parameter space due to large $t$ and $b$ quark Yukawa 
couplings 
and large $\st$- and $\sb$-mixings, and that this could have a 
decisive impact on $H^+$ searches at future colliders. 
In this paper we extend this study to the decay of the neutral Higgs 
bosons $H^0$ and $A^0$.

The lighter stop $\st_1$ can be much lighter than the other 
squarks and even lighter than the $t$ quark due to large 
$\st_L$-$\st_R$ mixing being proportional to the large top 
Yukawa coupling $h_t$ and the $\st$-mixing parameters $A_t$ 
and $\mu$ \cite{7}. 
Similarly, the lighter sbottom $\sb_1$ can also be much 
lighter than the other squarks \cite{5}. 
In the case of large $\st$- and $\sb$- mixings one also expects 
the couplings of $H^0\st\st$, $A^0\st\st$, $H^0\sb\sb$ and 
$A^0\sb\sb$ to be large. 
Here we show explicitly 
that the modes $\st\bar{\st}$ and $\sb\bar{\sb}$ can 
indeed dominate the $H^0$ and $A^0$ decays in a wide range of 
the MSSM parameters.

First we summarize current experimental limits on the squark masses. 
The D$\emptyset$ group at FNAL \cite{8,9} obtained 
mass bounds for squarks and gluinos from 
the negative search for them. 
However, they obtained \underline{no} mass bound for the 
mass-degenerate five flavors of squarks (\sq) (other than the stop 
(\st)) within the MSSM in case the lightest neutralino $\sz_1$ 
(assumed to be the lightest SUSY particle (LSP)) is heavier than 
$\sim$ 70GeV ; this is mainly due to the massive LSP effect. 
On the other hand, the LEP-I experiments set the general bound on the 
\sq mass $m_\sq$ \nge  45GeV 
\cite{10,11}. 
Experiments dedicated to \st search at LEP1.5 \cite{11} and 
Tevatron \cite{12} have set some limits on the lighter stop 
($\st_1$) mass for relatively light $\sz_1$, e.g., typically 
$m_{\st_1} \nge 55$GeV (for no \st -mixing) \cite{11} and 
$m_{\st_1} \nge 100$GeV \cite{12}, respectively. 
However, they set \underline{no} limit on $m_{\st_1}$ for 
$m_{\sz_1} \nge 50$GeV \cite{11,12}.

In the MSSM the properties of the charginos $\sw^\pm_i$ 
( $i = 1, 2$) and  neutralinos $\sz_j$ ($j = 1, \cdots, 4$) are 
completely determined by the parameters $M$, $\mu$ and 
$\tan\beta = v_2 / v_1$, assuming 
$M' = (5/3) \tan^2\theta_W M$. 
Here $M$ ($M'$) is the SU(2) (U(1)) gaugino mass, $\mu$ is the 
higgsino mass parameter, and $v_1$ ($v_2$) is the vacuum 
expectation value of the Higgs $H^0_1$ ($H^0_2$) \cite{1}. 
Here $m_{\sw^+_1}$ $<$ $m_{\sw^+_2}$ and 
$m_{\sz_1}$ $<$ $\cdots$ $<$ $m_{\sz_4}$. 
To specify the squark sector additional (soft SUSY breaking) 
parameters $M_{\tilde{Q}}$, $M_{\tilde{U}}$, $M_{\tilde{D}}$ 
(for each generation) and $A$ (for each flavor) are necessary. 
The mass matrix for stops in the base of ($\st_L$, $\st_R$) reads 
\cite{7,3} 
\beq
{\cal M}_\st^2 = \left( \begin{array}{cc}m_{\st_L}^2 & a_t m_t \\ 
           a_t m_t & m_{\st_R}^2 \end{array}  \right)
\eeq
with 
\bea
m_{\st_L}^2 &=& M_{\tilde{Q}}^2+m_t^2+m_Z^2\cos 2\beta 
(I_t^3-e_t\sin^2\theta_W), \\ 
m_{\st_R}^2 &=& M_{\tilde{U}}^2+m_t^2+m_Z^2\cos 2\beta 
e_t\sin^2\theta_W,  \mbox{\quad and }\\ 
a_t m_t &=& m_t (A_t - \mu \cot\beta). 
\eea
For the \sb system analogous formulae hold but with 
$M^2_{\tilde{U}}$ replaced by $M^2_{\tilde{D}}$ in eq.(3), 
and instead of eq.(4), 
$a_b m_b = m_b (A_b - \mu \tan\beta)$.
$\sb_L$-$\sb_R$ mixing may be important for large 
$A_b$, $\mu$ and $\tan\beta$. 
The slepton sector is fixed by adding (soft SUSY breaking) 
parameters $M_{\tilde{L}}$ and $M_{\tilde{E}}$.

The masses and couplings of the Higgs bosons $H^\pm$, $H^0$, 
$h^0$ and $A^0$, including leading Yukawa corrections, are fixed by 
$m_{A^0}$, $\tan\beta$, $m_t$, $M_{\tilde{Q}}$, $M_{\tilde{U}}$, 
$M_{\tilde{D}}$, $A_t$, $A_b$ and $\mu$. 
$H^0$ ($h^0$) and $A^0$ are the heavier (lighter) CP-even and 
CP-odd neutral Higgs bosons, respectively. 
For the Yukawa corrections to the $h^0$ and $H^0$  masses and 
their mixing angle $\alpha$ we use the formulae of ref.\cite{13}. 
We include also the potentially large Yukawa corrections to the triple 
Higgs vertices of $H^0 h^0 h^0$, $H^0 A^0 A^0$ and $h^0 A^0 A^0$ 
using the formulae of ref.\cite{14}.

In the following, we take for simplicity 
$M_{\tilde{Q}} = M_{\tilde{U}} = M_{\tilde{D}}$ 
(for the third generation), 
$M_{\tilde{L}} = M_{\tilde{E}} = M_{\tilde{Q}}$, and 
$A_t = A_b = A_\tau \equiv A$. 
Thus we have as free parameters 
$m_{A^0}$, $M$, $\mu$, $\tan\beta$, $M_{\tilde{Q}}$ 
and $A$.

We calculate the widths of all possibly important 
modes of $H^0$ and $A^0$ decays: 
({\romannumeral 1}) 
$H^0$ $\to$ $t\bar{t}$, $b\bar{b}$, $c\bar{c}$, 
$\tau^-\tau^+$, $W^+W^-$, $Z^0Z^0$, $h^0h^0$, $A^0A^0$, 
$W^\pm H^\mp$, $Z^0A^0$, $\st_i\bar{\st}_j$, 
$\sb_i\bar{\sb}_j$, $\sl^-_i\sl^+_j$, $\sn_\ell\bar{\sn}_\ell$ 
($\ell = e, \mu, \tau$), $\sw^+_i\sw^-_j$, 
$\sz_k\sz_l$, and ({\romannumeral 2}) 
$A^0$ $\to$ $t\bar{t}$, $b\bar{b}$, $c\bar{c}$, 
$\tau^-\tau^+$, $Z^0h^0$, $\st_1\bar{\st}_2$, $\st_2\bar{\st}_1$, 
$\sb_1\bar{\sb}_2$, $\sb_2\bar{\sb}_1$, 
$\stau^-_1\stau^+_2$, $\stau^-_2\stau^+_1$, $\sw^+_i\sw^-_j$, 
$\sz_k\sz_l$. 
Formulae for these widths are found in ref.\cite{2}. 
In principle, also the decays 
$H^0 \to \sq_\alpha\bar{\sq}_\alpha$ 
($q = u, d, c, s$ and $\alpha = L, R$) could contribute 
via their gauge couplings. 
As the squarks of the first two generations are supposed 
to be heavy, these decays will be strongly phase-space 
suppressed. 
Even if they were kinematically allowed, they would have a 
rate at most comparable to that of 
$H^0$ $\to$ $\sl^-_i\sl^+_j$ and $\sn_\ell\bar{\sn}_\ell$ 
(see fig.~2 below).
We neglect loop induced decay modes (such as $H^0 \to gg$ 
and $\gamma\gamma$) and three-body decay modes \cite{15}.

In order not to vary many parameters, in the following we 
fix $m_t=180$GeV and $\mu=300$GeV, and take the values of 
$M$ and $\tan\beta$ such that $m_{\sz_1} \simeq 70$GeV for 
which the D$\emptyset$ bounds on $m_\sq$ and $m_{\st_1}$ \cite{8,12} 
and the LEP1.5 bound on $m_{\st_1}$ \cite{11} disappear.

In fig.1 the contour lines for the branching ratios 
$B(H \to \st\bar{\st}, \sb\bar{\sb})$ $\equiv$ \\
$\sum_{i,j=1,2}(B(H\to\st_i\bar{\st}_j)+
B(H\to\sb_i\bar{\sb}_j))$ 
($H$ $=$ $H^0$ or $A^0$) are plotted in the 
$A$ -- $M_{\tilde{Q}}$ plane for 
($m_{A^0}$(GeV), $M$(GeV), $\tan\beta$) $=$ 
($450$, $160$, $2$)(a), 
($500$, $160$, $2$)(b), 
($450$, $146$, $12$)(c) and 
($500$, $146$, $12$)(d), for which 
$m_{\sw^+_1}$(GeV) $=$ $128$, $128$, $131$ and $131$, 
respectively. 
In the plots we have required $m_{\st_1, \sb_1, \sl}$ 
$>$ $m_{\sz_1}$ ($\simeq 70$GeV). 
In the allowed regions of fig.1 we find 
$m_{H^0}$ $\simeq$ $m_{A^0}$ and $m_{h^0} > 55$GeV, 
which satisfies the LEP limit $m_{h^0} > 44$GeV \cite{16}. 
We see that the branching ratios 
$B(H^0\to\st\bar{\st}, \sb\bar{\sb})$ and 
$B(A^0\to\st\bar{\st}, \sb\bar{\sb})$ can be larger than 
50\% in a sizable region. 
In this region, these SUSY decay modes dominate over
the conventional modes.

In fig.2 we show the $m_{A^0}$ dependence 
(in the $m_{A^0}$ range for $m_{h^0} > 50$GeV) 
of the important branching ratios of $H^0$ and $A^0$ decays 
for ($M_{\tilde{Q}}$(GeV), $A$(GeV), $M$(GeV), $\tan\beta$) $=$ 
($102$, $325$, $160$, $2$) (a, b), 
($218$, $405$, $145$, $30$) (c), and  
($145$, $274$, $146$, $12$) (d).  
In these three cases we have 
($m_{\st_1}$, $m_{\st_2}$, $m_{\sb_1}$, $m_{\sb_2}$, $m_{\sw^+_1}$) 
$=$ ($100$, $270$, $100$, $115$, $128$) GeV (a, b), 
($80$, $386$, $80$, $304$, $132$) GeV (c), and 
($80$, $310$, $80$, $199$, $131$) GeV (d). 
We see that in these cases the sum of the $\st\bar{\st}$ and 
$\sb\bar{\sb}$ modes dominates the $H^0$ and $A^0$ decays in a 
wide range of $m_{A^0}$. 
Here note that $m_{H^0}$ $\simeq$ $m_{A^0}$ in the $m_{A^0}$ 
range shown here, and that the $A^0$ does not couple to 
$\st_i\bar{\st}_i$, $\sb_i\bar{\sb}_i$ and 
$\stau^+_i\stau^-_i$ ($i=1,2$). 

As for $h^0$ decay, we have found that the decay 
$h^0 \to \st_1\bar{\st}_1$ is kinematically allowed 
only in a very limited region of the MSSM parameter space. 

The reason for the dominance of the $\st\bar{\st}$ and 
$\sb\bar{\sb}$ modes in the $H^0$ decay is as follows : 
The modes $t\bar{t}$, $b\bar{b}$, $\st\bar{\st}$ and 
$\sb\bar{\sb}$ (whose couplings to $H^0$ are essentially 
$\sim h_t \sin\alpha$, $\sim h_b \cos\alpha$, 
$\sim (A - \mu \cot\alpha) h_t \sin\alpha$ and 
$\sim (A - \mu \tan\alpha) h_b \cos\alpha$, respectively) 
can be strongly enhanced relative to the other modes 
due to the large Yukawa couplings $h_{t,b}$. 
In addition, the $\st\bar{\st}$ and $\sb\bar{\sb}$ modes 
can dominate over the $t\bar{t}$ and 
$b\bar{b}$ modes, respectively, in the case the $\sq$-mixing 
parameters $A$ and $\mu$ are large. 
Furthermore, in this case $\st_1$ and $\sb_1$ tend to be light 
due to a large mass-splitting. 
Here note that the effects of the bottom Yukawa coupling 
$h_b$ and $\sb$- mixing play a very important role for 
large $\tan\beta$ (see fig.2c). 
The reason for the dominance of the $\st\bar{\st}$ and 
$\sb\bar{\sb}$ modes in the $A^0$ decay is similar to 
that in the $H^0$ decay.

Quite generally, $B(H\to\st\bar{\st}, \sb\bar{\sb})$ depends 
on the parameters $m_{A^0}$, $M_{\tilde{Q}}$, $A$, $\mu$, 
$\tan\beta$ and more weakly on $M$. 
For a given $m_{A^0}$ the strongest dependence is that on 
$M_{\tilde{Q}}$ to which $m_\st$ and $m_\sb$ are sensitive 
(see fig.1). 
$B(H\to\st\bar{\st}, \sb\bar{\sb})$ can be quite large 
in a substantial part of the parameter region kinematically 
allowed for the $\st\bar{\st}$ and $\sb\bar{\sb}$ modes. 
We find that the dominance of the $\st\bar{\st}$ and 
$\sb\bar{\sb}$ modes is fairly insensitive to the assumption 
$M_{\tilde{Q}}$ $=$ $M_{\tilde{U}}$ $=$ $M_{\tilde{D}}$ 
$=$ $M_{\tilde{L}}$ $=$ $M_{\tilde{E}}$. 
As seen in fig.1 the dependence on $A$ is also strong. 
Concerning the assumption $A_t=A_b=A_\tau$, we have 
found no significant change of 
$B(H\to\st\bar{\st}, \sb\bar{\sb})$ ($H=H^0, A^0$) as 
compared to fig.2, when we take 
$A_{b,\tau}/A_t$ $=$ $\pm 0.5$, $\pm 1$, $\pm 2$ keeping 
$A_t=A$. 
As $|\mu|$ increases, the dominance of the 
$\st\bar{\st}$ and $\sb\bar{\sb}$ modes becomes more 
pronounced because of the increase of the 
($H^0$, $A^0$) couplings to $\st\bar{\st}$ and $\sb\bar{\sb}$. 
We also find that 
$B(H\to\st\bar{\st}, \sb\bar{\sb})$) ($H=H^0,A^0$) 
is nearly invariant under ($\mu$, $A$) $\to$ ($-\mu$,$-A$).

Here we have not included QCD radiative corrections to the 
hadronic modes. 
It is shown in ref.\cite{17} that both the standard QCD 
correction (due to gluon-quark loop) and SUSY QCD 
correction (due to gluino-squark loop) to the widths 
of $H^0$, $A^0$ $\to$ $t\bar{t}$, $b\bar{b}$ can be 
large, but that the two corrections can partly or even totally cancel 
each other. The QCD corrections to 
$H^0$, $A^0$ $\to$ $\st\bar{\st}$, $\sb\bar{\sb}$ 
are not known. It is shown in ref.\cite{6} that 
the QCD corrections to the closely related process 
$H^+ \to \st\bar{\sb}$ are significant, but that 
they do not invalidate the tree-level conclusion 
of ref.\cite{5} on the dominance of the $\st\bar{\sb}$ 
mode in a wide parameter region. 
As the structures of the couplings of $H^0$ and $A^0$ 
to $\st\bar{\st}$ and $\sb\bar{\sb}$ ($t\bar{t}$ and $b\bar{b}$) 
are similar to those of $H^+$ to 
$\st\bar{\sb}$ ($t\bar{b}$), we can expect that the QCD 
corrections 
would not invalidate our tree-level conclusion on the 
dominance of the $\st\bar{\st}$ and $\sb\bar{\sb}$ 
modes in a large parameter region.

As for the signatures of the $H^0$ and $A^0$ decays, 
typical $\st\bar{\st}$ and $\sb\bar{\sb}$ signals are shown in table~1.  
They have to be compared with the conventional $t\bar{t}$ and 
$b\bar{b}$ signals, respectively, 
($H^0$, $A^0$) $\to$ $t\bar{t}$ $\to$ ($W^+b$)($W^-\bar{b}$) $\to$ 
($f\bar{f'}$)$b$($f\bar{f'}$)$\bar{b}$ (i.e., 6 jets (j's), 4j's $+$ 1 
isolated
 charged lepton 
($\ell^{\pm}$) $+$ missing energy-momentum ($\misp$), or 
2j's $+$ $\ell^+\ell^{'-}$ $+$ $\misp$) and 
($H^0$, $A^0$) $\to$ $b\bar{b}$ (i.e.,  2j's). 
Note that $B(\st_1\to c\sz_1)$ $\simeq$ $1$ if 
$m_{\st_1}$ $<$ $m_{\sw^+_1, \sl, \sn, \sb_1}$ 
and 
$m_{\sz_1}$ $<$ $ m_{\st_1}$ $<$ $m_{\sz_1}+m_{t, W}$ 
(in cases (a), (b) and (d) of table~1), and $B(\st_1\to c\sz_1)$ $\simeq$ 
$0$ otherwise (in case (c)) \cite{18}. 
(In principle there is a region
$m_b+m_W+m_{\sz_1} < m_{\st_1} < m_b+m_{\sw^+_1}$ 
where also the decay $\st_1 \to bW^+\sz_1$ 
plays a role \cite{bwc}. 
However, this decay does not occur in the 
cases considered here.)
As seen in table~1, the $\st\bar{\st}$ ($\sb\bar{\sb}$) signals have 
general 
features which distinguish them from the $t\bar{t}$ ($b\bar{b}$) signals : 
({\romannumeral 1}) more $\misp$ due to the emission of two LSP's and 
hence less energy-momentum of jets and the isolated-charged-lepton in 
case of a 
short decay chain, or ({\romannumeral 2}) more jets and/or more 
isolated-charged-leptons in case of a larger decay chain. 
Moreover, depending on the values of the MSSM parameters, 
the $\st\bar{\st}$ and $\sb\bar{\sb}$ signals could have the following 
remarkable features : 
({\romannumeral 1}) production of same-sign dileptons 
$\ell^{\pm}\ell^{'\pm}$ (e.g., in case (g)), which 
could yield same-sign isolated dilepton events such as 
$e^+e^-$ $\to$ $H^0Z^0$ $\to$ ($\ell^+\ell^{'+}$ or 
$\ell^-\ell^{'-}$) $+$ j's $+$ $\misp$ ; 
({\romannumeral 2}) less bottom-jet activity (e.g., in cases 
(a), (b), (d) with ($h^0$, $Z^0$) $\to$ 
($\ell^+\ell^-$ or $\nu\bar{\nu}$) ) or more 
bottom-jet activity (e.g., in cases (c), (d), (f) with 
($h^0$, $Z^{(*)}$) $\to$ $b\bar{b}$) ;  
and ({\romannumeral 3})  emission of a real $Z^0$ or $h^0$ 
(e.g., in cases (b), (c), (d), (f)). 
The identification of the sign of charged leptons and the tagging of 
$b$- and $c$-quark jets, $h^0$, $Z^0$ and $W^\pm$ would be 
very useful \cite{2} in discriminating the $\st\bar{\st}$ and 
$\sb\bar{\sb}$ signals from the $t\bar{t}$  and $b\bar{b}$ signals 
as well as in suppressing the background.

The suitable places for $H^0$ and $A^0$ search would be 
$e^+e^-$ colliders \cite{19} and hadron supercolliders \cite{4}. 
If the $\st\bar{\st}$ and $\sb\bar{\sb}$ modes dominate the $H^0$ and $A^0$ 
decays, it decisively influences the signatures of $H^0$ and $A^0$. 
For example, it can strongly suppress the conventional discovery 
modes of $H^0$ and $A^0$ at the hadron supercollider \cite{4}, 
such as the "gold-plated" $H^0 \to Z^0Z^0$ mode 
(viable for small $\tan\beta$ and $2m_Z$ $\nle$ $m_{H^0}$ $\nle$ $2m_t$) 
(see fig.2a) and $H^0$, $A^0$ $\to$ $\tau^-\tau^+$ mode 
(viable for large $\tan\beta$) (see figs.2c, d). 
Clearly it would be necessary to perform a detailed Monte-Carlo study 
to separate the signals from the background. 
Such a study is, however, beyond the scope of this article.

We have shown that the SUSY modes $\st\bar{\st}$  and $\sb\bar{\sb}$ can 
dominate the $H^0$ and $A^0$ decays in a large allowed region of the MSSM 
parameter space due to large $t$ and $b$ quark Yukawa couplings and 
large \st- and \sb-mixings. The $\st\bar{\st}$  and $\sb\bar{\sb}$ modes 
have 
very distinctive signatures as compared to the conventional modes 
$t\bar{t}$, $b\bar{b}$ and $\tau^-\tau^+$. 
This could decisively influence the $H^0$ and $A^0$ search at future 
$e^+e^-$ and hadron colliders.

While preparing this manuscript a paper by Djouadi et al. 
\cite{20} appeared dealing with a similar subject. 
They also point out the possible importance of the squark pair modes 
in $H^0$ and $A^0$ decays. 
They, however, studied them in a strongly constrained supergravity model 
in contrast to our work, which is in the general 
framework of the MSSM.

\section*{Acknowledgements}
The authors thank Prof. V. Barger for a valuable communication 
on ref.\cite{14}. 
The work of A.B., H.E., and W.M. was supported by the ``Fonds zur 
F\"orderung der 
wissenschaftlichen Forschung'' of Austria, project no. P10843-PHY.
The work of Y. Y. was supported in part by the U. S.~Department of Energy 
under 
Grant No.~DE-FG02-95ER40896 and in part by the University of Wisconsin 
Research 
Committee with funds granted by the Wisconsin Alumni Research Foundation. 
The work of T. K. was supported in part by 
the Grant-in-Aid for Scientific Research Program 
from the Ministry of Education, Science and Culture of Japan, No.08640388.

\vfill\eject

\newpage
\pagestyle{plain}
\setcounter{page}{15}
\vspace{20mm}
\section*{Figure Captions}
\renewcommand{\labelenumi}{Fig.\arabic{enumi}} \begin{enumerate}

\vspace{6mm}
\item
Contour lines of 
$B$($H^0\to\st\bar{\st}$, $\sb\bar{\sb}$) (a, c) and 
$B$($A^0\to\st\bar{\st}$, $\sb\bar{\sb}$) (b, d) in the 
$A$ -- $M_{\tilde{Q}}$ plane for 
($m_{A^0}$(GeV), $M$(GeV), $\mu$(GeV), $\tan\beta$) $=$ 
($450$, $160$, $300$, $2$)(a), 
($500$, $160$, $300$, $2$)(b), 
($450$, $146$, $300$, $12$)(c) and 
($500$, $146$, $300$, $12$)(d). 
The contour of $B = 0$ (dashed line) represents the 
kinematical boundary ; the lines ({\romannumeral 3}), 
({\romannumeral 4}) and ({\romannumeral 5}) 
are the contour lines of 
$2m_{\st_1} = m_{H^0}$ ($\simeq 450$ GeV), 
$2m_{\sb_1} = m_{H^0}$ ($\simeq 450$ GeV) and 
$m_{\sb_1}+m_{\sb_2} = m_{A^0}$ ($= 500$ GeV), 
respectively. 
The shaded area is excluded by the requirement 
$m_{\st_1, \sb_1, \sl}$ $>$ $m_{\sz_1}$ 
($\simeq 70$ GeV) ; 
the lines ({\romannumeral 1}) and ({\romannumeral 2}) are 
the contour lines of $m_{\st_1}$ and $m_{\sb_1}$ $=$ $70$ GeV, 
respectively, above which $m_{\st_1}$ and $m_{\sb_1}$ $>$ $70$ GeV. 

\vspace{6mm}
\item
The $m_{A^0}$ dependence of important branching ratios 
of the $H^0$ and $A^0$ decays for 
($M_{\tilde{Q}}$(GeV), $A$(GeV), $M$(GeV), $\mu$(GeV), $\tan\beta$) $=$ 
($102$, $325$, $160$, $300$, $2$) (a, b), 
($218$, $405$, $145$, $300$, $30$) (c) and  
($145$, $274$, $146$, $300$, $12$) (d).  
The sum over all mass eigenstates and/or flavors is taken for 
$B(\st\bar{\st})$, $B(\sb\bar{\sb})$, 
$B(\sw^+\sw^-)$, $B(\sz\sz)$, $B(\stau^+\stau^-)$ and 
$B(\sn\bar{\sn})$. 
The shoulders in the curves for the $\st\bar{\st}$ and 
$\sb\bar{\sb}$ modes correspond to opening of the 
$\st_2\bar{\st}_2$ and $\sb_2\bar{\sb}_2$ channels, respectively. 
\end{enumerate}


\clearpage
\begin{thebibliography}{99}
\bibitem{1}
H. E. Haber and G. L. Kane, Phys. Rep. \underline{117} (1985) 75. 
\bibitem{2}
J. F. Gunion, H. E. Haber, G. L. Kane, and S. Dawson, 
The Higgs Hunter's Guide, Addison-Wesley (1990) 
\bibitem{3}
J. F. Gunion and H. E. Haber, Nucl. Phys. \underline{B272} 
(1986) 1; \underline{B402} (1993) 567 (E).
\bibitem{4}
Z. Kunszt and F. Zwirner, Nucl. Phys. \underline{B385} (1992) 3. 
\bibitem{5}
A. Bartl, K. Hidaka, Y. Kizukuri, T. Kon and W. Majerotto, 
Phys. Lett. \underline{B315} (1993) 360. 
\bibitem{6}
A. Bartl, H. Eberl, K. Hidaka, T. Kon, W. Majerotto and 
Y. Yamada, Phys. Lett. \underline{B373} (1996) 117. 
\bibitem{7}
J. Ellis and S. Rudaz, Phys. Lett. \underline{B128} (1983) 248. 
\bibitem{8}
D$\emptyset$ Collab., S. Abachi et al., Phys. Rev. Lett. 
\underline{75} (1995) 618.
\bibitem{9}
E. Gallas, 
Proc. of the XXXIst Rencontres de Moriond, "QCD and High Energy Hadronic 
Interactions", Les Arcs, Savoie, France, 23-30 March, 1996, 
(FERMILAB-Conf-96/114-E) ; J. Conway, 
talk at "SUSY96", Maryland, May 1996. 
\bibitem{10}
J. F. Grivaz, 
Proc. of the XXVIIIth Rencontres de Moriond on Electroweak interactions 
and unified theories, (Les Arcs, Savoie, France, March, 1993). 
\bibitem{11}
S. Asai, Proc. of the Lake Louise Winter Institute 
"Topics in Electroweak Physics", Alberta, 18-24 Feb. 1996 
(ICEPP preprint UT-ICEPP 96-01) ; 
ALEPH Collab., D. Buskulic et al., Phys. Lett. \underline{B373} (1996) 
246 ; 
L3 Collab., H. Nowak and A. Sopczak, L3 Note 1887, 1996 ; 
OPAL Collab., S. Asai and S. Komamiya, OPAL Physics Note PN-205, 1996 ; 
DELPHI Collab., DELPHI 96-101, 1996. 
\bibitem{12}
D$\emptyset$ Collab., S. Abachi et al., Phys. Rev. Lett. 
\underline{76} (1996) 2222.
\bibitem{13}
J. Ellis, G. Ridolfi and F. Zwirner, Phys. Lett. 
\underline{B262} (1991) 477. 
\bibitem{14}
V. Barger, M. S. Berger, A. L. Stange and R. J. N. Phillips, 
Phys. Rev. \underline{D45} (1992) 4128.
\bibitem{15}
A. Djouadi, J. Kalinowski and P. M. Zerwas, 
Z. Phys. \underline{C70} (1996) 435. 
\bibitem{16}
Review of Particle Properties, 
Phys. Rev. \underline{D54} (1996) 1 ; 
DELPHI Collab., P. Abreu et al., Z. Phys. \underline{C67} (1995) 69. 
\bibitem{17}
J. A. Coarasa, R. A. Jim\'enez and J. Sol\`a, 
hep-ph/9511402 ; see also R. A. Jim\'enez and J. Sol\`a, 
hep-ph/9511292, and A. Bartl, H. Eberl, K. Hidaka, T. Kon, 
W. Majerotto and Y. Yamada,  Phys. Lett. \underline{B378} (1996) 167.
\bibitem{18}
I. I. Bigi and S. Rudaz, Phys. Lett. \underline{B153} (1985) 335 ; 
K. I. Hikasa and M. Kobayashi, Phys. Rev. \underline{D36} (1987) 724. 
\bibitem{bwc}
W. Porod and T. Woehrmann, Vienna prprint UWThPh-1996-45. 
\bibitem{19}
A. Djouadi, J. Kalinowski and P. M. Zerwas, and P. Janot, 
Proceedings "$e^+e^-$ Collisions at 500GeV : The Physics Potential", 
Munich - Annecy - Hamburg 1991, ed. P. M. Zerwas, 
DESY report 'DESY 92 -123A' ; 
A. Djouadi, J. Kalinowski and P. M. Zerwas, Z. Phys. \underline{C57} (1993) 569. 
\bibitem{20}
A. Djouadi, J. Kalinowski, P. Ohmann and P. M. Zerwas, 
preprint hep-ph/9605339. 
\end{thebibliography}
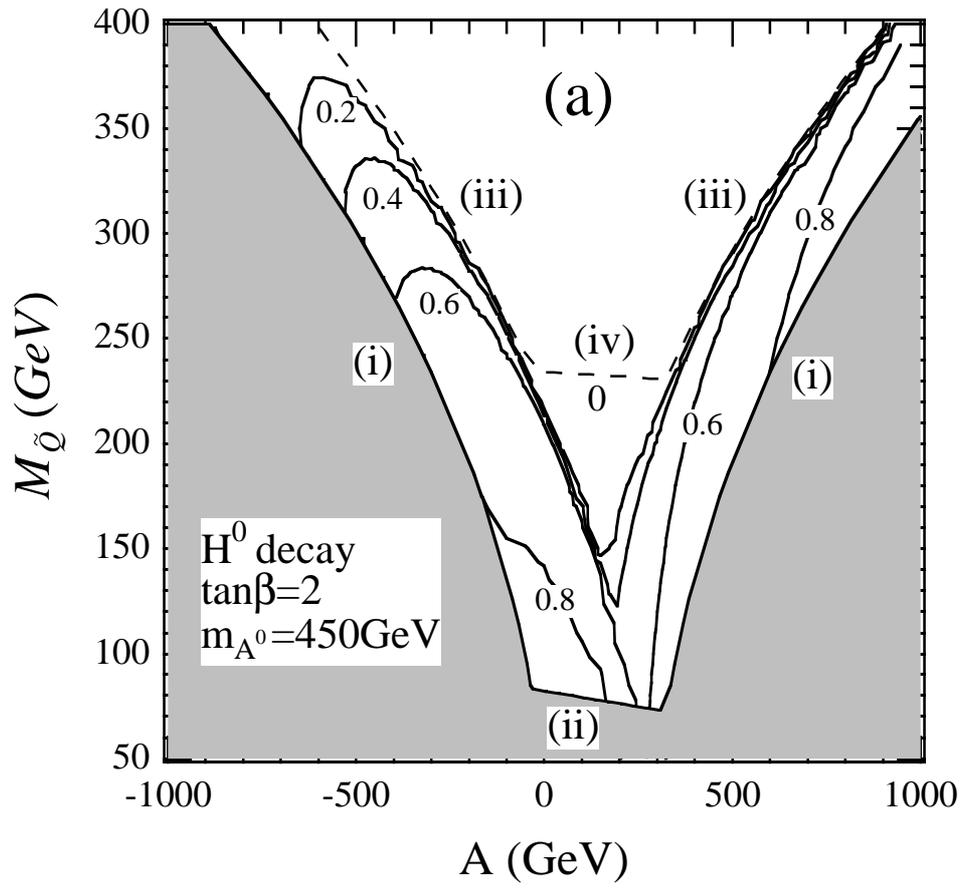
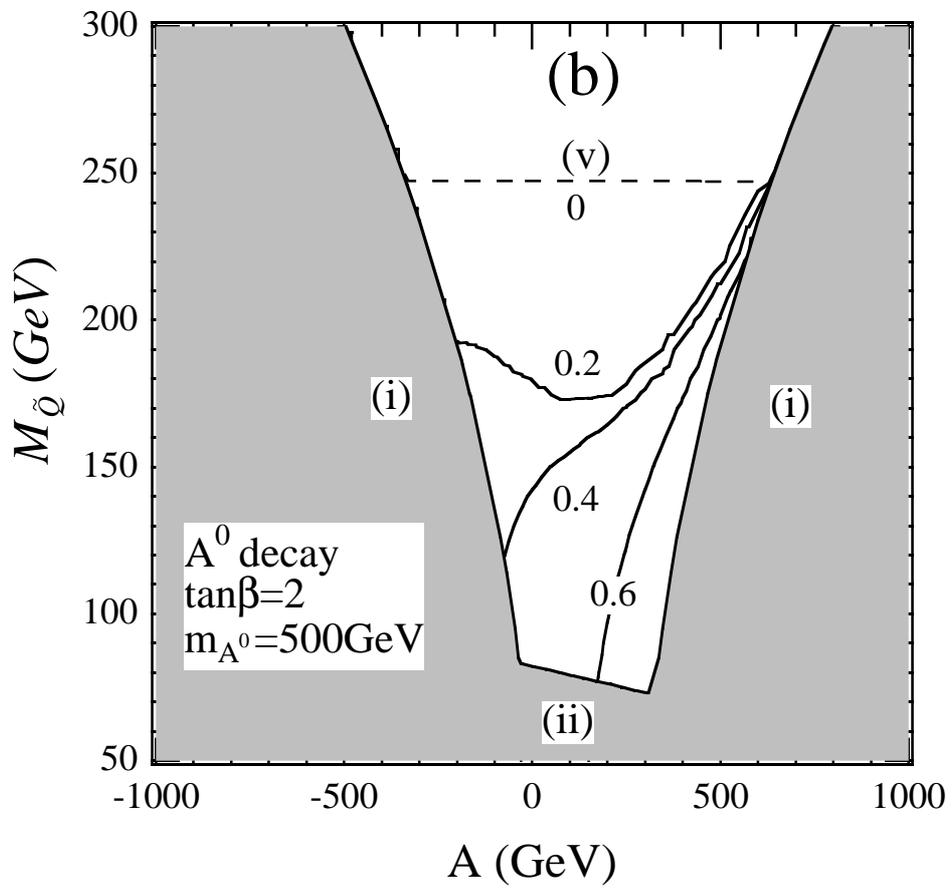
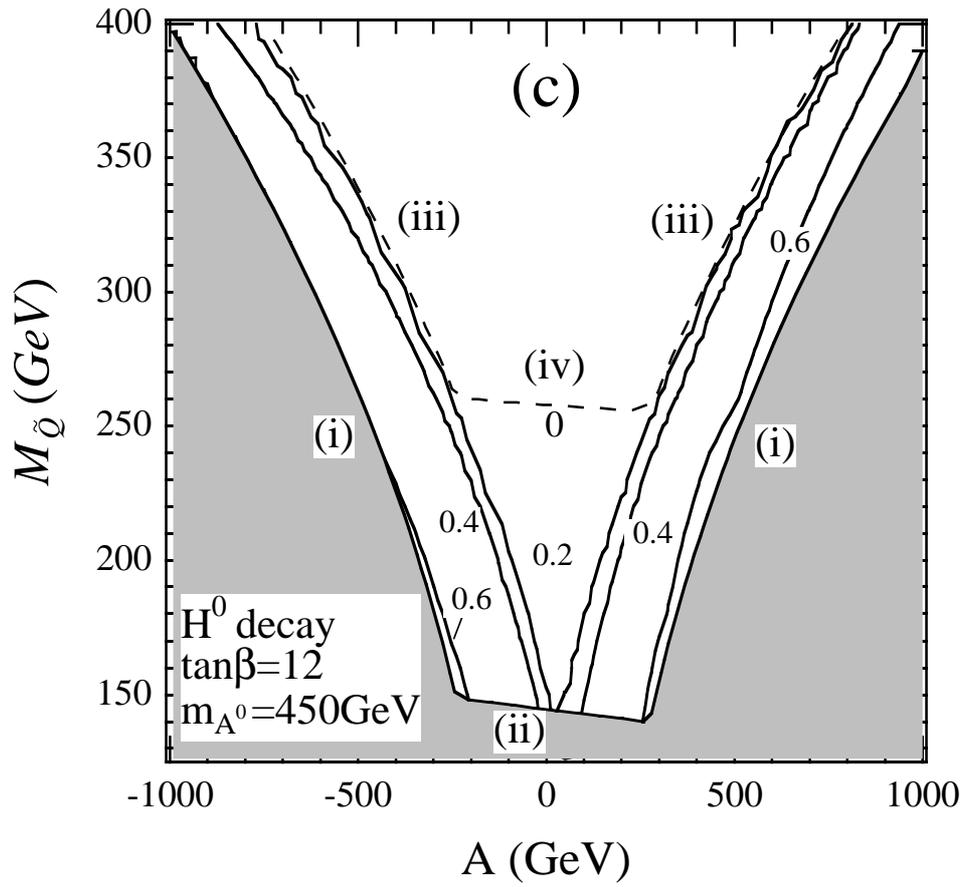
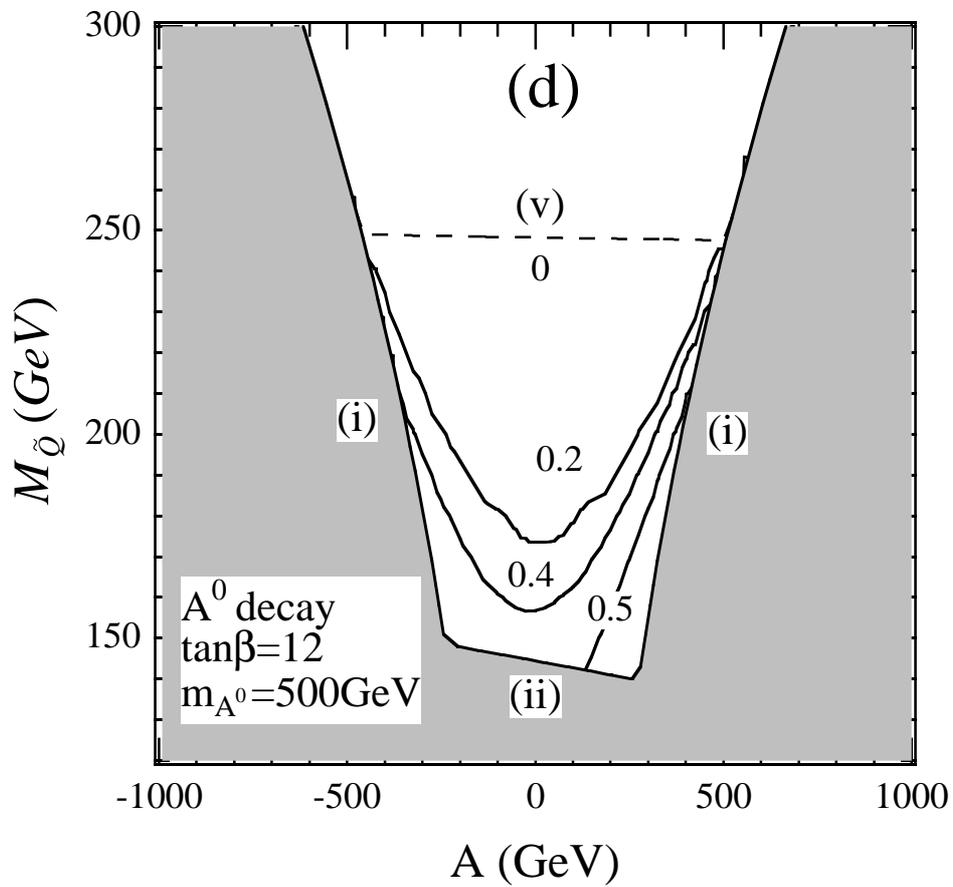
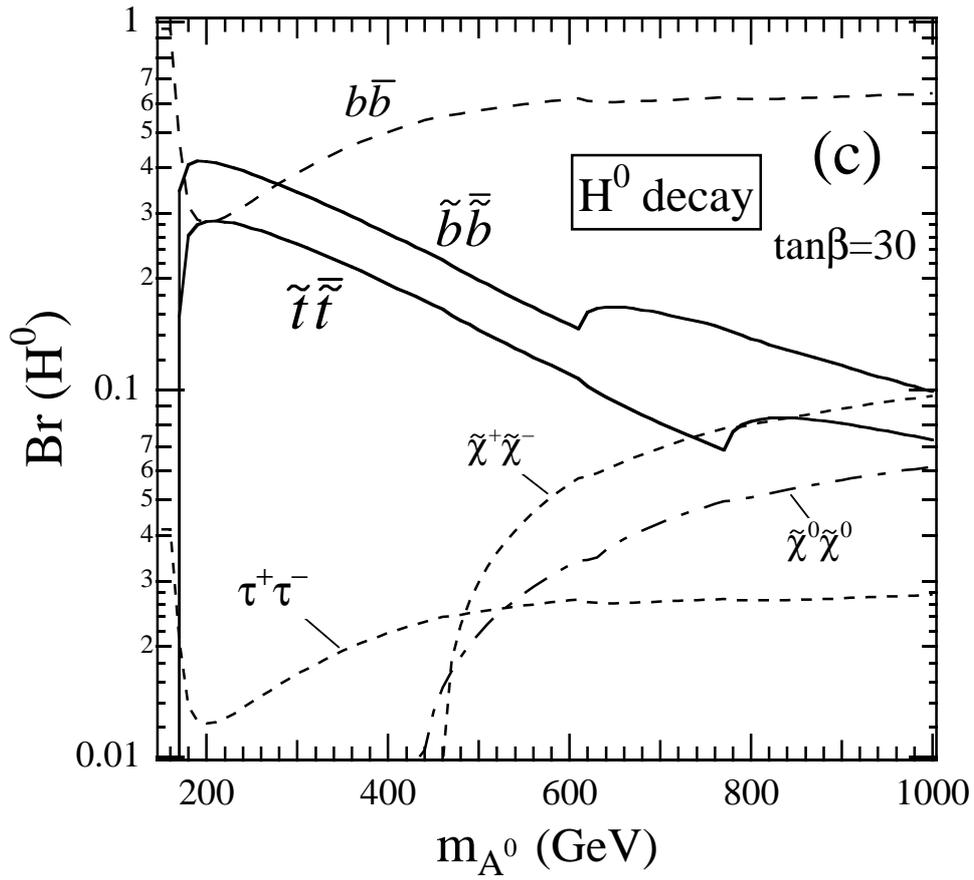
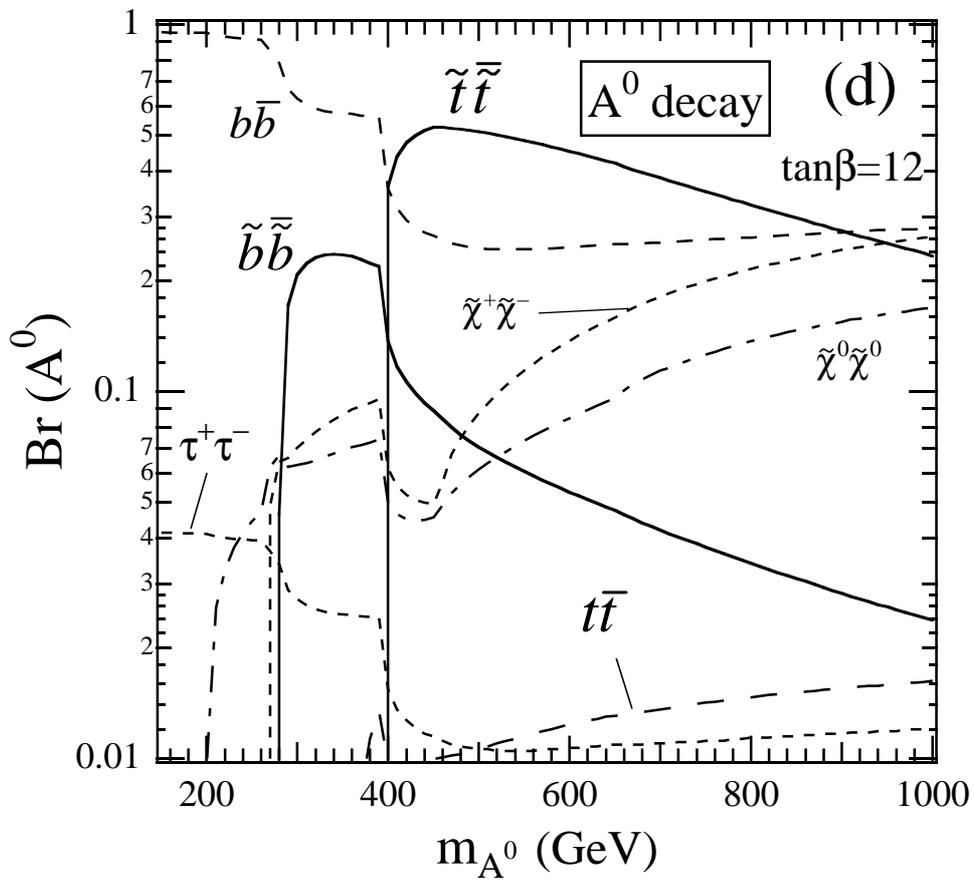
\end{document}